\documentstyle{elsart}

\input epsf

\begin{document}
\begin{frontmatter}
\title{
A fast programmable trigger for isolated cluster counting
in the BELLE experiment 
}
\author{H.J.Kim\thanksref{LSU},}
\author{S.K.Kim, S.H.Lee, T.W.Hur\thanksref{ss},}
\author{C.H.Kim, F.Wang\thanksref{inhp},}
\author{I.C.Park\thanksref{lg},}
\address{Department of Physics, Seoul National University, Seoul 151-742, Korea}

\author{Hee-Jong Kim,}
\address{Department of Physics, Yonsei University, Seoul 120-749, Korea}

\author{B.G.Cheon,}
\address{KEK, Tsukuba, Ibaraki 305-0801, Japan}

\author{E. Won\thanksref{address}\thanksref{kek}.}
\address{Research Institute for Basic Sciences, Seoul National University, Seoul 151-742, Korea}

\thanks[LSU] {Also affiliated with Department of Physics and Astronomy, Louisiana State University, Baton Rouge, LA 70803, USA}
\thanks[ss] {Permanent Address : SsangYong Information \& Communications Corp., 24-1 Jeo-dong 2-ga, Jung-gu, Seoul 100-748, Korea}
\thanks[inhp] {Permanent Address : Institute of High Energy Physics Academia Sinica, China}
\thanks[lg] {Permanent Address : LG Semicon Co.,Ltd. 1, Hyangjeong-dong, Hungduk-gu, Cheongju-si 361-480, Korea}
\thanks[address] {Corresponding author; E-mail: eiwon@bmail.kek.jp; Tel: +81 298 64 5351; FAX: +81 298 64 2580}
\thanks[kek] {Also affiliated with KEK, Tsukuba, Ibaraki 305-0801, Japan}

\begin{abstract}
 We have developed a fast programmable trigger processor board based on a
field programmable gate array  and a complex programmable
logic device for use in the BELLE experiment.  
The trigger board accommodates 144 ECL input signals, 2 NIM input 
signals, 24 ECL output signals, and 
the VME bus specification.  An asynchronous
trigger logic for counting isolated clusters is used.
We have obtained trigger latency of 50 ns with a full access to  input 
and output signals via a VME interface. The trigger logic can be modified
at any time depending on the experimental conditions.
\end{abstract}


\end{frontmatter}

\section{Introduction}
Fast, complex, general-purpose trigger systems are required for modern
particle physics experiments.  Although custom-made CMOS gate arrays are
used for extremely fast applications such as first-level triggers 
($\sim$ 25 ns)
for LHC experiments\cite{lhc}, 
field programmable gate arrays (FPGAs) are an attractive option for
environments that require a less demanding speed ($<$ 100 ns) but
a more flexible
trigger logic implementation.  The logic of FPGA-based trigger systems
can be readily changed as the nature of signal
and background conditions vary.   Such trigger systems are
flexible and can be adapted to many different applications. 
Commercial products that have these functionalities exist 
(for example, the Lecroy 2366 Universal Logic Module, Lecroy Co.)
and can be used for implementing  rather simple trigger logic. In the
case of the
calorimeter trigger for the BELLE 
experiment, the number of channels, data transfer rates, 
and the complexity
of the trigger logic preclude the use of commericially
available devices.  We developed a 9U VME\cite{vme} 
module  that accommodates
more than a hundred ECL signals for the triggering purpose. The resulting
board is a general purpose asynchronous programmable trigger board that 
satisfies VME specifications.

\section{Trigger requirements for the BELLE Experiment}
\begin{figure}
\begin{center}
\epsfxsize=5.0in
\epsfbox{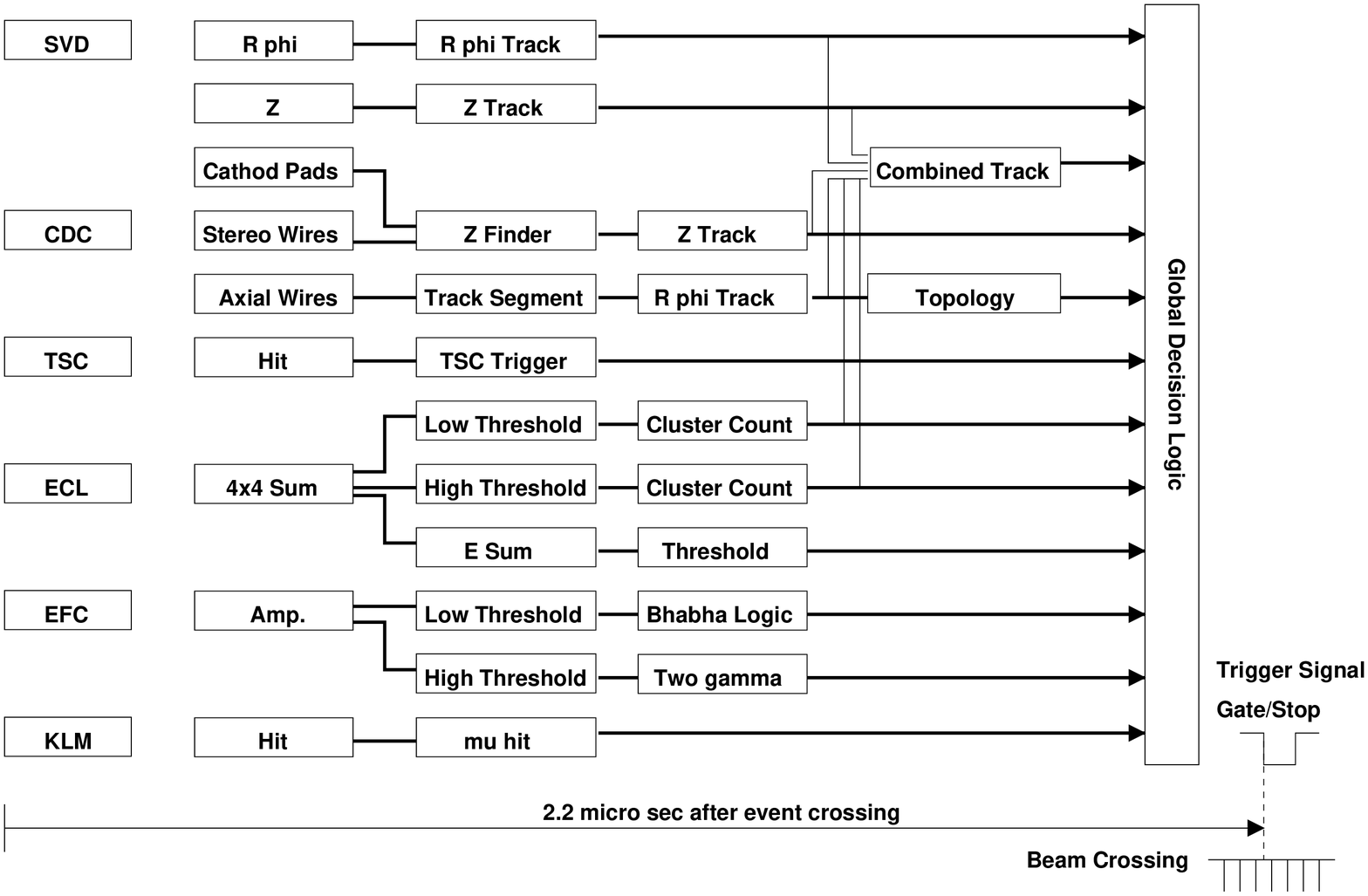}
\caption
{A block diagram of the BELLE trigger system. Information from all sub 
detectors is gathered in the GDL, where the trigger decision is made.}
\protect\label{trig}
\end{center}
\end{figure}

The BELLE experiment\cite{belle} 
at KEK in Japan, is designed to exloit the physics potential
of KEKB, a high luminosity, asymmetric $e^+ e^-$ collider operating at 
a cm energy (10.55 GeV) corresponding to the
$\Upsilon(4S)$ resonance.  In particular, BELLE is designed to test the 
Kobayashi-Maskawa mechanism for CP violation in B meson sector. The 
KEKB design luminosity is 1 $\times$ 
$10^{34}$cm$^{-2}$s$^{-1}$ with  a bunch crossing rate of 2 ns.  
The BELLE detector consists of seven subsystems; 
a silicon vertex detector (SVD), 
a central drift chamber (CDC), 
an aerogel Cherenkov counter (ACC), an array of
trigger and time of flight scintillation counters (TOF/TSC), 
an electro-magnetic calorimeter (ECL), 
$K_L$ and muon detectors (KLM) and 
extreme forward calorimeters (EFC).  A
1.5 Tesla axial magnetic field is produced by a superconducting
solenoid located outside of the ECL. 
The KLM is outside of the solenoid and provides
a return yoke for the detector's magnetic field. 
The BELLE trigger system requires logic
with a level of sophistication that can distinguish and select desired
events from a large number of 
background processes that may change depending on the conditions of the
KEKB storage ring system.
Figure~\ref{trig} shows a schematic view of the 
BELLE trigger system.
As shown in Fig.~\ref{trig}, 
the trigger information from individual detector components
is formed in parallel and combined in one final stage. This scheme 
facilitates the formation of redundant triggers that rely either only on
information from the calorimeter or from the tracking systems. 
The final event trigger time is determined 
by requiring a coincidence between the beam-crossing RF signal and the output of the 
final trigger decision logic. The timing and width of the subsystem
trigger signals are adjusted so that their outputs always cover
the beam-crossing at a well defined fixed delay of 2.2 $\mu$s
from the actual event crossing. 

The ECL is a highly segmented array of $\sim$ 9000 CsI(Tl) crystals
with silicon photodiode
readout installed inside the coil of the solenoid magnet. Preamplifier outputs
from each crystal are added in summing modules located just outside of the
BELLE detector and then split into two streams with two different shaping 
times (1 $\mu$s and 200 ns): 
the slower one for the total energy measurement
and the faster one for the trigger. 
For the trigger,
signals from a group of crystals 
are summed to form a trigger cell (TC), 
discriminated, digitized (as differential ECL logic signals),
and fed into five Cluster
Counting Modules (CCMs) that count
the number of isolated clusters in the calorimeter. 
In total, the ECL has 512 trigger cells: 432 in the barrel region and 80 in the
endcaps.
The trigger latency of the CCM trigger 
board is $\sim$ 150 ns.  Each module 
accepts 132 inputs and outputs 16 logic signals. 
(The actual board can accommodate a maximum of
144 inputs and provide as many as 24 output signals;  for BELLE  
we have chosen to use 132 input and 16 output lines 
per board). 

Given the complexity discussed above and the required flexibility, we 
chose to use a complex FPGA  to apply
the isolated clustering algorithm and a CPLD device 
in order to match the VME bus specifications.  For the FPGA, we use 
an XC5215-PG299  chip that has 484 configurable logical blocks 
(CLBs), and for the CPLD, an XC95216-5PQ160, which provides 4,800 usable gates. 
Once the CPLD is loaded, it permanently holds 
all of the VME bus specification logic.
In contrast, the trigger logic contained in the FPGA is lost during 
a power down, and must be reconfigured during start-up, 
either from an on-board PROM or from a 
computer (VME master module)
through VME bus.  This takes a few milliseconds. 
In the following we describe in some
detail the trigger logic design of the CCM board
and how we achieve our performance requirements.

\section{Logic Design}
\begin{figure}
\begin{center}
\epsfxsize=5.0in
\epsfbox{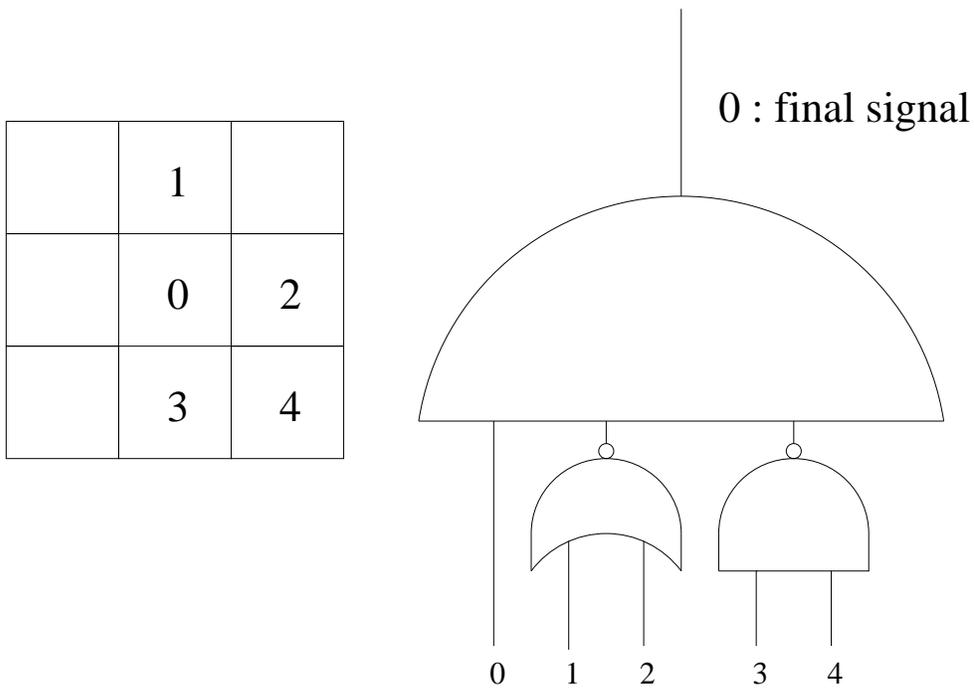}
\caption
{A logic diagram for the isolated cluster number counting. Among nine 
trigger cells,
only the three center cells and center and bottom right cells are considered
in isolated cluster counting logic.}
\protect\label{icn}
\end{center}
\end{figure}

\begin{figure}
\begin{center}
\epsfxsize=5.0in
\epsfbox{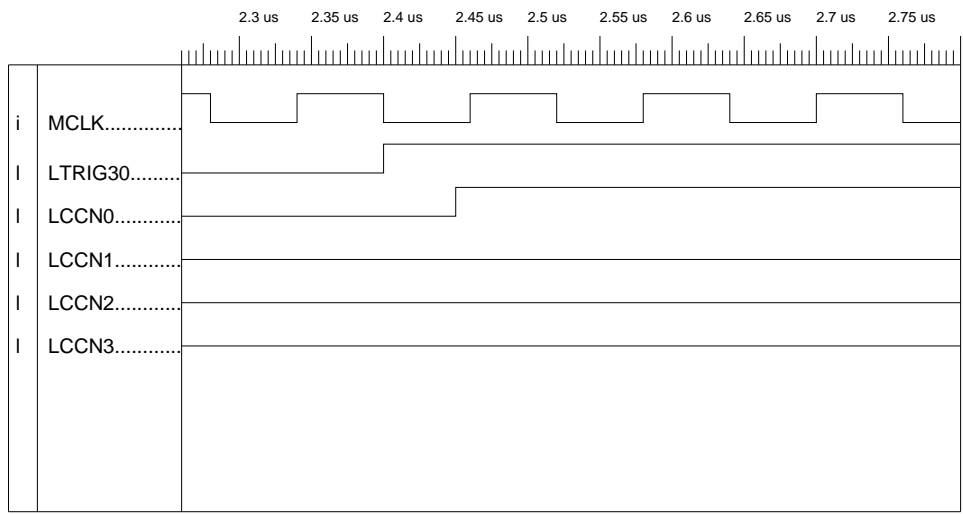}
\caption
{The simulated timing for ICN counting. From the top, the system clock
 (MCLK),
the input
trigger signal (LTRIG30), and the four ICN bits (lowest to highest, LCCN0-3) are 
shown. The time difference between LTRIG30 and LCCN0 is measured to
be 47.3 ns (One unit on the top is 5 ns).}
\protect\label{timing}
\end{center}
\end{figure}

We  used XACT$^{\sc TM}$
software provided by Xilinx\cite{xilinx} 
to design, debug and
simulate our logic. The trigger processor board accepts the differential
ECL logic signals from the
calorimeter trigger cells. There are many possible strategies 
for finding and counting
the number of isolated clusters (ICN) among
the calorimeter trigger cells.  But, since the 
trigger decision has to be made within a limited time period, 
a simple algorithm is desirable.  We devised simple logic that counts only
one cluster from a group of connected clusters. For
the case of a number of connected clusters,  we count only 
{\it the upper most cluster in the right most column} among them. This is
demonstrated for a 3 $\times$ 3 trigger cell array  in 
Fig.~\ref{icn}.  Here, the trigger cell under the counting operation 
is numbered as ``0''.  If the cell ``0'' satisfies the logic diagram
shown in Fig.~\ref{icn}, it is considered to be a single isolated cluster. We
have applied this simple logic to the output of GEANT-based\cite{geant} 
full Monte Carlo simulation
of various $B$ decay modes as well as Bhabha 
scattering events and compared the
perfect cluster number and the cluster number returned by the above logic.
The results are summarized in Table~\ref{icnsim}.
In all the cases, the discrepancies between the perfect cluster counting
and the isolated cluster counting logic are below the 1 \% level;
despite its simplicity, the counting logic works 
exceptionally well. 
This simple clustering logic is applied to over 132 input signals and the
number of isolated clusters are then tallied.  In addition to 
the cluster counting logic, we also  delay the
132 input and 16 output signals and register them in a set of FIFO RAMs 
(the pattern register) located on the board.  The 
signals are delayed (in order for them to be correctly phased) by approximately 
800 ns by means of an 8 MHz delay pulse and
stored in FIFO RAMs at the trigger decision. 
The delay time can be easily changed by
modifying the logic.   The pattern register allows a continuous
monitoring of the operation of 
the CCM module.  The recorded cluster and ICN bits are read out through
the VME bus independently of the ICN counting system. 
The FPGA counts the number of clusters asynchronously and 
the simulated timing diagram in Fig.~\ref{timing} indicates that the time
needed for the ICN counting is  47 ns.

 In order to satisfy the complete VME bus specification, a set of logical
blocks (Address Decoder, Interrupter, Control Logic, Configuration
Control Logic, CSR, and FIFO RAM Control)
are developed and downloaded into the CPLD. The logical blocks are
designed as a VME A24/D32 slave interface.
Comparators are used to
decode addresses being probed by the master module. Status 
bits are implemented in order to check the status of the configuration
completion of FPGA chip and triggering process itself. Control bits are
implemented to stop the output of the triggering signal, 
to start the output of the triggering signal, 
to enable the reconfiguration of the FPGA chip via a PROM or the VME bus,
and to control the FIFO RAM that serves as the pattern register.
All the functionalities were tested  extensively  during
the development phase and  completely debugged before they were
implemented in the experiment.

\section{Hardware Implementation}
\begin{figure}
\begin{center}
\epsfxsize=5.0in
\epsfbox{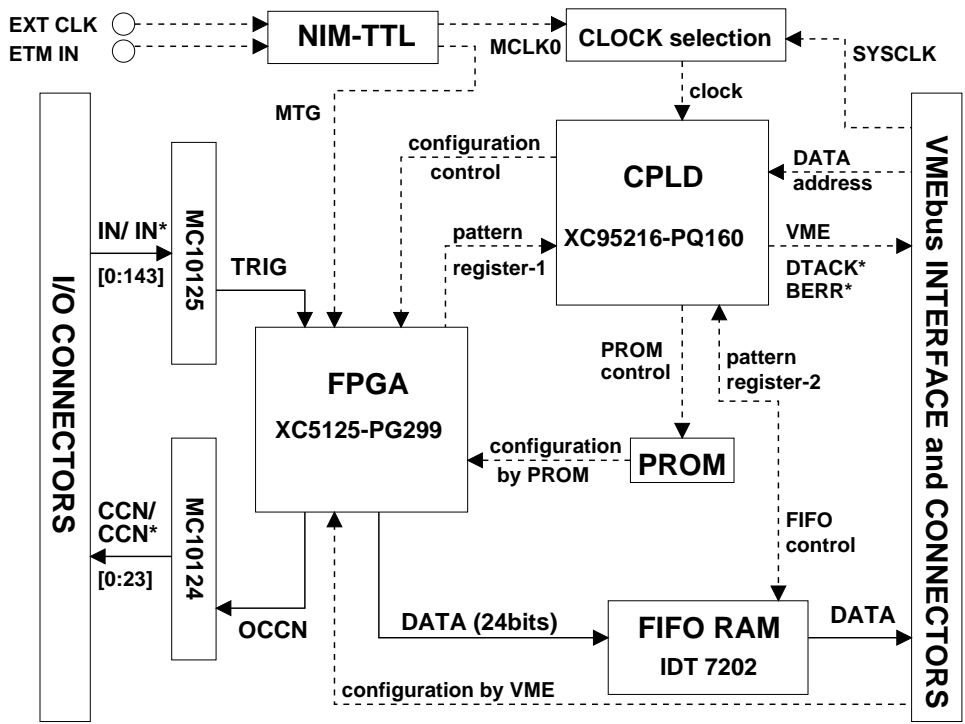}
\caption
{A simplified schematic of the CCM trigger board.}
\protect\label{block}
\end{center}
\end{figure}

\begin{figure}
\begin{center}
\epsfxsize=5.0in
\epsfbox{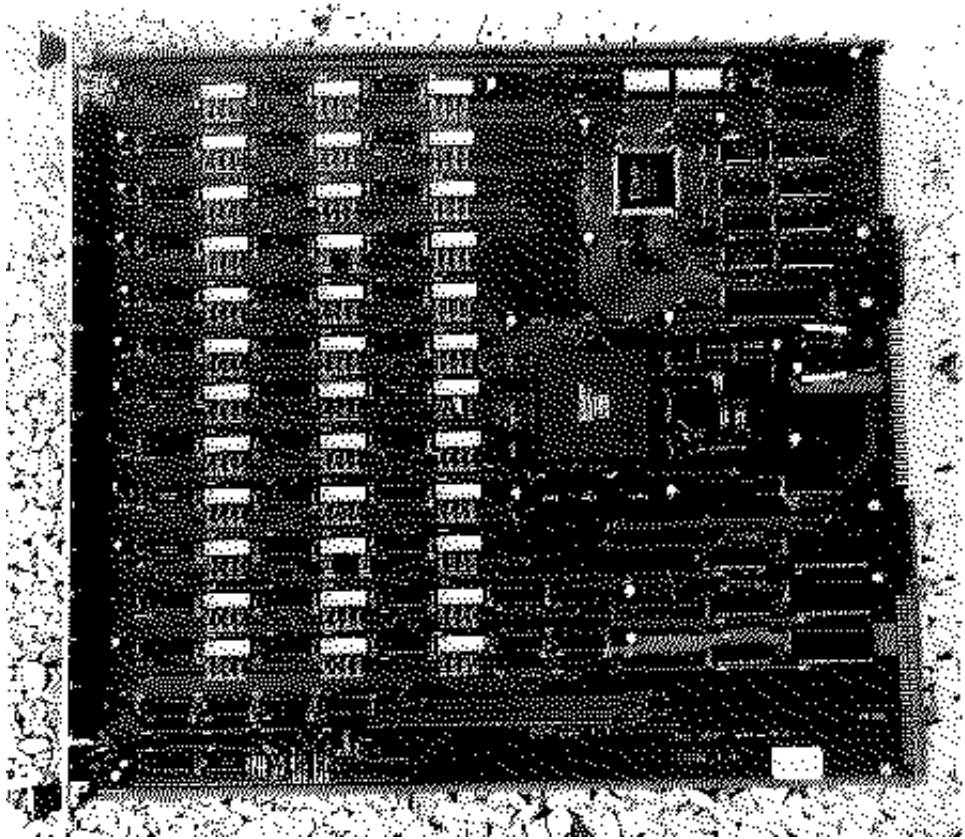}
\caption
{An assembled CCM trigger board. The FPGA chip is the big 
square-shaped chip near the center (gray color) and the CPLD is smaller square
surface-mounted chip near the top (black color).}
\protect\label{photo}
\end{center}
\end{figure}

The CCM module houses the main FPGA chip for the ICN counting, the
CPLD chip for implementing the VME bus specifications, ECL-TTL and NIM-TTL
converters, the PROM holding the FPGA configuration, and the
FIFO RAM pattern register. 
A schematic diagram  and an 
 assembled board are shown in Figs.~\ref{block} and ~\ref{photo},
respectively.
The printed circuit board is a VME 9U size four-layer board. All
connectors, switches, components, and downloading circuitry
are mounted on one side of the board. 
The logic signals to and from the FPGA are TTL CMOS, and are 
interfaced with the differential ECL logic signals to the rest of
the trigger and data acquisition system.  Standard 10124 (10125) chips
with 390 $\Omega$ pull down resisters (56 $\times$ 2 $\Omega$ termination
resisters) are used to convert TTL to ECL (ECL to TTL). The input polarity
is such that a positive ECL edge produces a negative TTL edge at the FPGA
input.  Also on-board are 
several discrete-component, NIM-TTL converters that interface with two external
NIM control signals: the master trigger signal (MTG) and the external clock.
Three 7202 CMOS asynchronous FIFO chips ( 3 $\times$ 1024 Bytes )
provide the pattern register. 
The actual registration for one event
includes 132 inputs, 16 outputs, 8 reserved bits,
10 memory address bits, and 2 unsed bits; 
a total of 146 bits are registered in the three
FIFO chips.

Programs for the FPGA chip can be downloaded from
an on-board PROM (Master Serial Mode)
or via the VME bus (Peripheral Asynchronous Mode). 
We use an XC17256D Serial Configuration PROM
and the clustering logic
is downloaded by a PROM writer that is controlled by a personal computer.
The choice of the VME master module 
is the FORCE\cite{force} SUN5V, 
a 6U VME bus CPU board that has a 110 MHz microSPARC-II
processor running Solaris 2.5.1.  Accessing the CCM from the VME
master module is simply done by mapping the device (in our case, the CCM) into 
the memory of the master module.  From there, the 
clustering logic can also be loaded into
the FPGA chip. 
All of the control software was
developed in this master module with GNU\cite{gnu} gcc and g++ 
compilers. An object-oriented graphical user interface based
on the  ROOT\cite{root} framework was also developed. 
Resetting the module, downloading the logic to FPGA
from the PROM or the VME bus, and the FIFO reading are all implemented in the 
graphical user interface.
Programs for the CPLD chip are downloaded through an on-board connector
from the parallel port of a personal 
computer and it enables the downloading
of the CPLD program  whenever necessary. 

The base address of the board is set by a 8-pin dip switch on board. 
A hardware reset switch that resets the FPGA, the CPLD, and the  FIFO RAMs is 
provided on the front panel. 
There are four LEDs indicating power on/off, 
MTG in, and two configuration of FPGA completion (LDC 
and SX1).  Two fuses (250V 2A) and four capacitors 
(100 $\mu$F) are on $\pm$ 5 V lines for the protection purpose.

\begin{figure}
\begin{center}
\epsfxsize=5.0in
\epsfbox{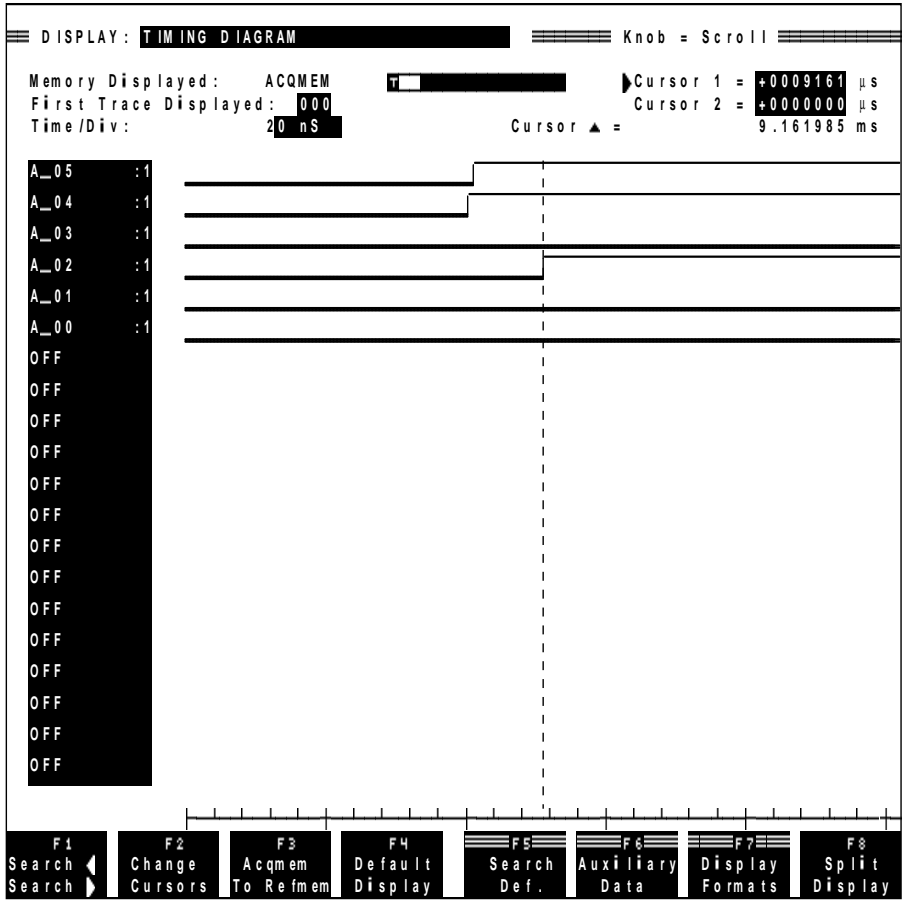}
\caption
{Measured timing results for ICN counting. Starting from the top,
two input ECL signals and
output ICN bits (lowest to highest bits) are displayed.  Each time division
is 20 ns.}
\label{timeresult}
\protect\end{center}
\end{figure}

The trigger board has been fully tested and the results have been compared 
with software simulations.  Test results are shown in Fig.~\ref{timeresult},
where a cluster-counting time of approximately 50 ns is found, which is
in good agreement with the 47 ns time predicted by the simulation. 

\section{Performance with e$^+$e$^-$ collisions}
 The BELLE detector started taking e$^+$e$^-$ collision data
with all subsystems, the data 
acquisition systems, and accompanying
trigger modules operational in early June of 1999.  Six
CCM modules installed in the electronics hut counted isolated
clusters from the e$^+$e$^-$ collision in the 
calorimeter.  Five CCM modules were used to
count isolated clusters from the five sections of the calorimeter;
the sixth module 
collected and summed the outputs from the other five.
The flexibility inherent in the design of the board allowed
the use some of the input and ouput
channels of the sixth module to generate triggers for Bhabha events
as well as calorimeter timing signals. 

\begin{figure}
\begin{center}
\epsfxsize=5.0in
\epsfbox{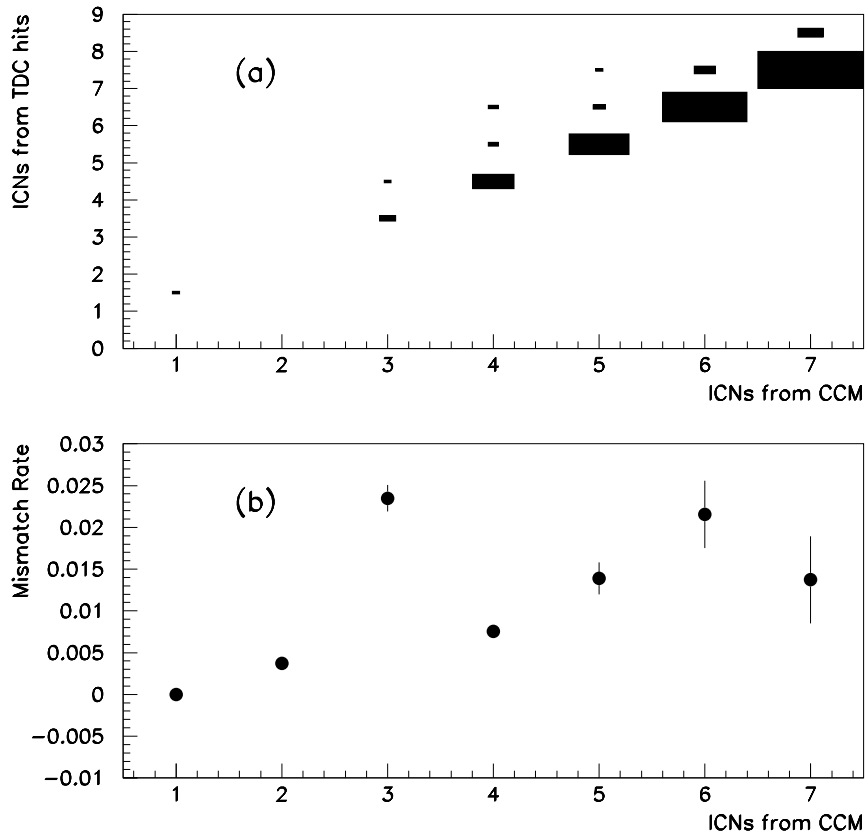}
\caption
{
The ICN-counting performace of the CCM modules. (a) ICN from TDC hit patterns 
vs. ICN from CCM modules, (b) mismatch rates between ICN from TDC hit patterns 
and from CCM as a function of the ICNs from CCM modules. 
}
\protect\label{rate}
\end{center}
\end{figure}

In a $\sim$ 100K event sample of actual triggers,
we found a  nearly perfect correspondence between the numbers
of isolated clusters provided by the trigger logic and
those inferred from TDC hit patterns that are available
at the offline analysis stage.
Figure~\ref{rate}(a) shows the correlation between the 
number of isolated clusters from TDC hit patterns
and ICN numbers from CCM modules. As is shown here, there are few
cases that ICN numbers from CCM modules are smaller than numbers from TDC hit
patterns.  Figure.~\ref{rate} (b) shows the 
mismatch rate between the TDC-based and-CCM based cluster numbers as 
a function of the TDC-based cluster numbers.
For more than 99.8 \% of the cases, the two numbers are
identical. We attribute the small level of inconsistency to the
limitations of the clustering counting logic 
(see section 3)
and the infrequent occurence of timing offset on the input signals.

\section{Conclusions}
We have developed a fast trigger processor board
utilizing FPGA and CPLD 
technology. It accommodates 144 ECL input signals and provides
24 ECL output signals. It  functions as a 9U VME module that
enables the loading of revised trigger logic and the online resetting 
of the module.
In addition, a pattern register on the board contains 
all of the  input/output ECL signals that were used in a process.
The isolated clustering logic is measured to have a time latency of
50 ns, in good agreement with the prediction
of the simulation.
Sufficient hardware
and software flexibility has been incorporated into the module to
make it well suited for dealing with a variety
of experimental conditions.

\begin{ack}
We would like to thank thr BELLE group for their  installation and maintenance
of the detector, and acknowledge support from KOSEF and
Ministry of Education (through BSRI) in Korea.
\end{ack}

\clearpage
\pagebreak


\begin{table}[ht]
\label{icnsim}
\begin{center}
\caption{The testing of the isolated cluster counting logic using simulations. 
The numbers in the first row indicate the difference between the 
perfect cluster number and the isolated cluster number from the logic.}
\vspace{0.4cm}
\begin{tabular}{|l|c|c|c|} \hline
ICN(logic)-ICN(perfect)   &  0 & 1 & 2\\ \hline\hline
Bhabha & 100 $\%$ & 0 $\%$ & 0 $\%$ \\ \hline
$B^0 \rightarrow J\/\psi K_S \rightarrow {l^+ l^-} \pi^+ \pi^-$    &
  99.15 $\%$ & 0.85 $\%$ & 0.00 $\%$ \\ \hline
$B^0 \rightarrow J\/\psi K_S \rightarrow {l^+ l^-} \pi^o \pi^o$    &
  98.55 $\%$ & 1.45 $\%$ & 0.00 $\%$ \\ \hline
$B^0 \rightarrow \pi^+ \pi^- $   &
  99.15 $\%$ & 0.85 $\%$ & 0.00 $\%$ \\ \hline
$B^0 \rightarrow \pi^o \pi^o $   &
  98.95 $\%$ & 1.05 $\%$ & 0.00 $\%$ \\ \hline
$B^+ \rightarrow \pi^+ \pi^o $   &
  98.95 $\%$ & 1.05 $\%$ & 0.00 $\%$ \\ \hline
$B^0 \rightarrow K^* \gamma  $ &
  99.15 $\%$ & 0.80 $\%$ & 0.05 $\%$ \\ \hline
$B^0 \rightarrow \rho^\pm \pi^\mp $   &
  99.15 $\%$ & 0.85 $\%$ & 0.00 $\%$ \\ \hline
\end{tabular}
\end{center}
\end{table}
\clearpage


\clearpage

\clearpage

\clearpage

\clearpage

\clearpage

\clearpage

\end{document}